\documentclass[vecphys]{svmult}

\usepackage{makeidx}         % allows index generation
\usepackage{graphicx}        % standard LaTeX graphics tool
\usepackage{multicol}        % used for the two-column index
\usepackage[bottom]{footmisc}% places footnotes at page bottom
\usepackage[latin1]{inputenc}
\usepackage[T1]{fontenc}
\usepackage[english]{babel}
\usepackage{epsfig}
\usepackage{amssymb}
\usepackage[centertags]{amsmath}
\usepackage{amsfonts}
\usepackage{makeidx}
\usepackage{graphics}
\usepackage{graphicx}
\usepackage{psfrag}
\usepackage{subfigure}
\usepackage{slashed}
\usepackage{dsfont}
\usepackage{eucal}
\usepackage{eufrak}
\usepackage{color}
\usepackage{bm}
\usepackage[square,comma,sort&compress]{natbib}
\usepackage{fancybox}
\usepackage{footnpag}
\usepackage{latexsym}
\usepackage{appendix}
\usepackage{fancyhdr}
\usepackage{textcomp}
\usepackage{latexsym}
\usepackage{slashed} % per inserire lo slash di Feynman
\usepackage{fancyhdr}
\usepackage{feynmf}

\def \g {\gamma}

\def\laq{~\raise 0.4ex\hbox{$<$}\kern -0.8em\lower 0.62
ex\hbox{$\sim$}~}
\def\gaq{~\raise 0.4ex\hbox{$>$}\kern -0.7em\lower 0.62
ex\hbox{$\sim$}~}

\begin{document}

\title*{Newly observed charmed states: the case of $X(3872)$}
\author{Stefano Nicotri \inst{}}


\institute{Universit\`a degli studi di Bari and INFN sezione di
Bari,\\via Orabona 4, I-70126, Bari, Italy\\
\texttt{stefano.nicotri@ba.infn.it} }

\maketitle

\begin{abstract}
  I briefly review the properties of the hidden charm
  meson $X(3872)$, discussing some puzzling aspects concerning its
  theoretical interpretation.
\end{abstract}

\section*{Introduction}
\label{sec:1}

Recently many new mesons have been observed in the open and hidden
charm sectors \cite{PDG,Reviews}. From the need of theoretical
interpretation of those states, the interesting possibility of
identifying new ``exotic'' structures has been pointed out. In the
following I discuss the case of $X(3872)$, which represents one of
the most interesting examples in this sense. In particular I focus
the attention on the properties of the radiative decays $X\to D\bar
D\g$ \cite{Colangelo:2007ph}.

% Always give a unique label
% and use \ref{<label>} for cross-references
% and \cite{<label>} for bibliographic references
% use \sectionmark{}
% to alter or adjust the section heading in the running head

\section*{$X(3872)$}
\label{sec:2}

\begin{figure}[t]
\begin{center}
\includegraphics[scale=0.7]{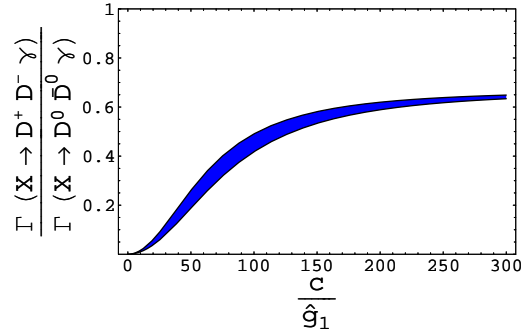}
\end{center}
\caption{\footnotesize{Ratio of charged $X \to D^+ D^- \gamma$ to
neutral $X \to D^0 \bar D^0 \gamma$ decay widths versus the ratio of
hadronic parameters $c/\hat g_1$.}}\label{fig:ratio} \vspace*{1cm}
\end{figure}

The experimental observations can be summarized as follows:
\begin{itemize}
\item the $X(3872)$  resonance has been found in $J/\psi \pi^+ \pi^-$ distribution by four experiments,
both in $B$ decays($B^{-(0)} \to K^{-(0)} X$), both in $p \bar p$
annihilation \cite{expX3872}. The  mass is $M=3871.9 \pm 0.6$ MeV
while the width remains unresolved:  $\Gamma < 2.3$ MeV (90 \% CL);
\item  there is no evidence of resonances  in the charged mode $J/\psi \pi^\pm\pi^0$ or in $J/\psi \eta$ \cite{babarkpi};
\item the state is not observed in $e^+ e^-$ annihilation;
\item  for $X$ produced in $B$ decays the ratio $\frac{B(B^0\to K^0 X)}{B(B^+\to K^+X)}=0.61 \pm 0.36 \pm 0.06$ is obtained \cite{babarkpi};
\item the dipion spectrum in $J/\psi \pi^+ \pi^-$ is peaked at large mass \cite{pipispectrum};
\item  the decay  in $J/\psi \pi^+ \pi^- \pi^0$ is observed \cite{belle3p} with $\frac{B(X\to J/\psi\pi^+\pi^-\pi^0)}{B(X\to J/\psi\pi^+\pi^-)}=1.0 \pm 0.4 \pm 0.3$:
this implies G-parity violation;
\item the radiative  mode $X \to J/\psi \gamma$ is found \cite{belle3p,babarpsigamma}  with
$\frac{B(X \to J/\psi \gamma)}{B(X \to J/\psi \pi^+ \pi^- )}=0.19
\pm 0.07$, therefore charge conjugation of the state is C=+1;
\item  the angular distribution of the final state is compatible with the spin-parity assignment $J^P=1^+$  \cite{Abe:2005iy};
\item  there is a  signal in $D^0 \bar D^0 \pi^0$ with  $\frac{B(X \to D^0 \bar D^0  \pi^0)}{B(X \to J/\psi \pi^+ \pi^- )}=9\pm4$ \cite{BelleDoDopi};
the near-threshold  $D^0 \bar D^0 \pi^0$  enhancement in $B \to D^0
\bar D^0 \pi^0 K$has a the peak at $M=3875.4 \pm 0.7 ^{+1.2}_{-2.0}$
MeV with $B(B\to K X \to K D^0 \bar D^0  \pi^0)=(1.27 \pm
0.31^{+0.22}_{-0.39}) \times 10^{-4}$ \cite{Gokhroo:2006bt} .
\end{itemize}
All the measurements are compatible with the assignment
$J^{PC}=1^{++}$, but if the $2^3P_1$ is identified with  $Y(3940)$,
another recently observed meson, there is an overpopulation of
$1^{++}$ $c \bar c$ mesons. This, together with the observation of
the G-parity violating $X$ transitions suggested the conjecture that
$X(3872)$ is not a charmonium $\bar c c$ state. Indeed, the
coincidence between the $X$ mass as averaged by PDG and the $D^{*0}
\overline D^0$ mass inspired the proposal that $X(3872)$ could be a
molecular quarkonium \cite{okun}, a $D^{*0}$ and $\overline D^0$
bound state with small binding energy due to a single pion exchange
\cite{molec}. In this scheme the wave function of $X(3872)$ is given
by \cite{voloshin1}:
\begin{equation}
|X(3872)>=a \, |D^{*0} \bar D^0+ \bar D^{*0}  D^0> + b \, |D^{*+}
D^-+  D^{*-}  D^+> + \dots
\end{equation}
(with $|b| \ll |a|$) and so one could explain why this state seems
not to have  definite isospin, why the mode $X \to J/\psi \pi^0
\pi^0$ was not  found, and why, if the binding mechanism is a single
pion exchange (which is in fact a controversial point), there are no
$D \overline D$ molecular states. The opposing remark concerning the
large value of the ratio $\frac{B(X \to J/\psi \pi^+ \pi^-
\pi^0)}{B(X \to J/\psi \pi^+ \pi^- )}$ is that the ratio of the
amplitudes is small: $ \frac{A(X \to J/\psi \rho^0)}{A(X \to J/\psi
\omega)}\simeq 0.2$, so that the isospin violating amplitude is 20\%
of the isospin conserving one, an effect that could be related to
the mass difference between neutral and charged $D$ mesons,
considering  the contribution of $DD^*$ intermediate states to $X$
decays \cite{suzuki,Meng:2007cx}. Moreover, the central value of the
mass measured in $D^0 \bar D^0 \pi^0$ is $4$ MeV higher than the PDG
value (with a large systematic error), a result difficult to explain
in the molecular picture \cite{Hanhart:2007yq}. It has also been
suggested that the molecular interpretation would imply that the
radiative decay in neutral $D$ mesons: $X \to D^0 \bar D^0 \gamma$
should be dominant with respect to $X \to D^+ D^- \gamma$
\cite{voloshin1}. However, assuming that $X(3872)$ is an ordinary
$J^{PC}=1^{++}$ charmonium and describing the $X(3872)\to D \bar D
\gamma$ amplitude by polar diagrams with $D^*$ and $\psi(3770)$ as
intermediate particles the ratio $R={\Gamma(X \to D^+ D^-
\gamma)\over \Gamma(X \to D^0 \overline D^0 \gamma)}$ depends only
on the ratio $c/\hat g_1$, $c$ being the parameter governing the
radiative $X\psi(3770)\g$ matrix element, and $\hat g_1$
representing the coupling $XD\bar D^*$ (or $XD^*\bar D$). The ratio
$R$ is plotted in fig.\ref{fig:ratio} versus ${c/\hat g_1}$ and it
is shown that in any case $R < 0.7$. Thus, there is a suppression of
the radiative $X$ decay mode into charged $D$ mesons with respect to
the mode with neutral $D$. Moreover, for small values of ${c/\hat
g_1}$ the ratio $R$ is tiny, so that this seems not a peculiar
feature of a molecular structure of $X(3872)$.

\begin{figure}[t]
\begin{center}
\includegraphics[scale=0.5]{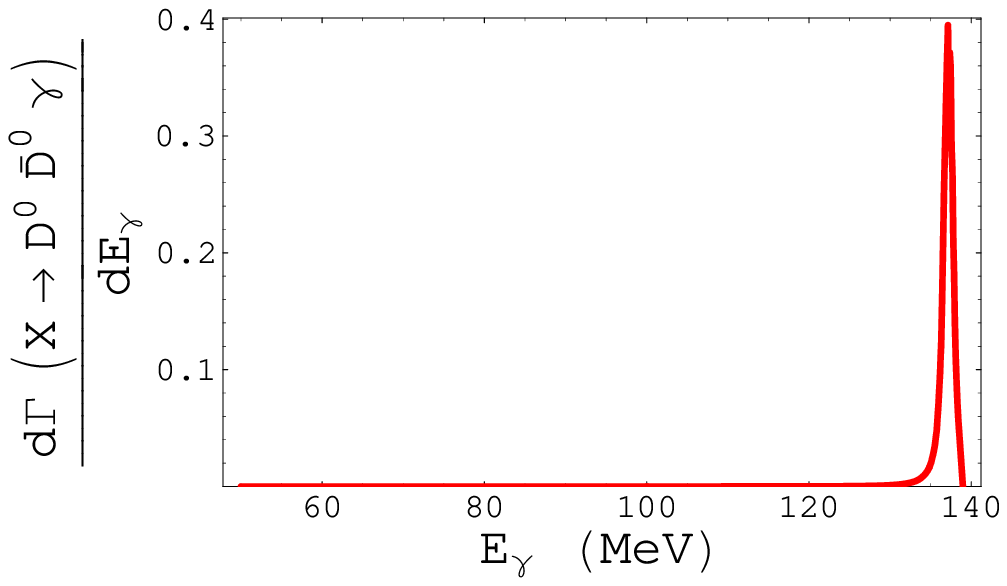} \hspace{0.35cm}
 \includegraphics[scale=0.5]{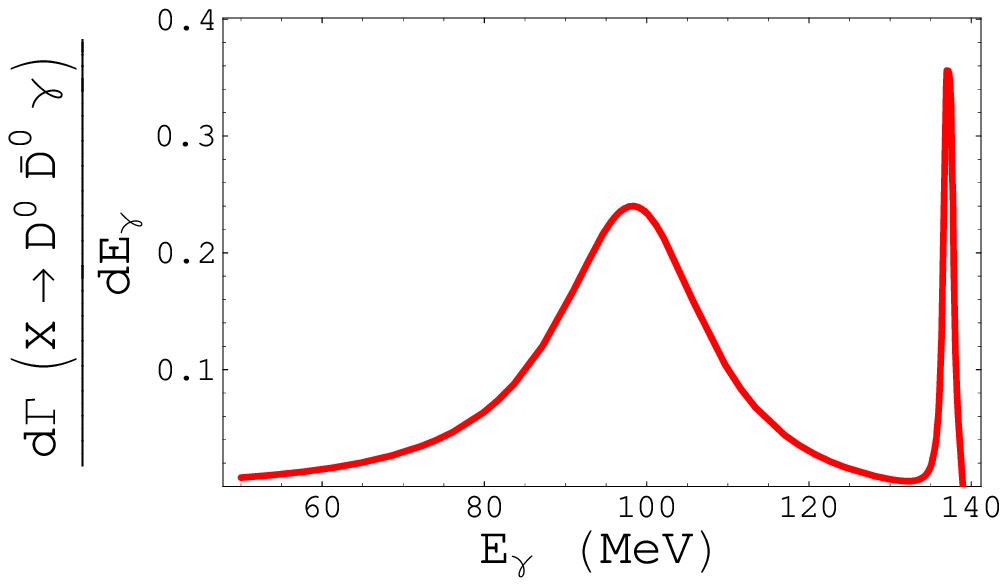}\\ \vspace{0.35cm}
 \includegraphics[scale=0.5]{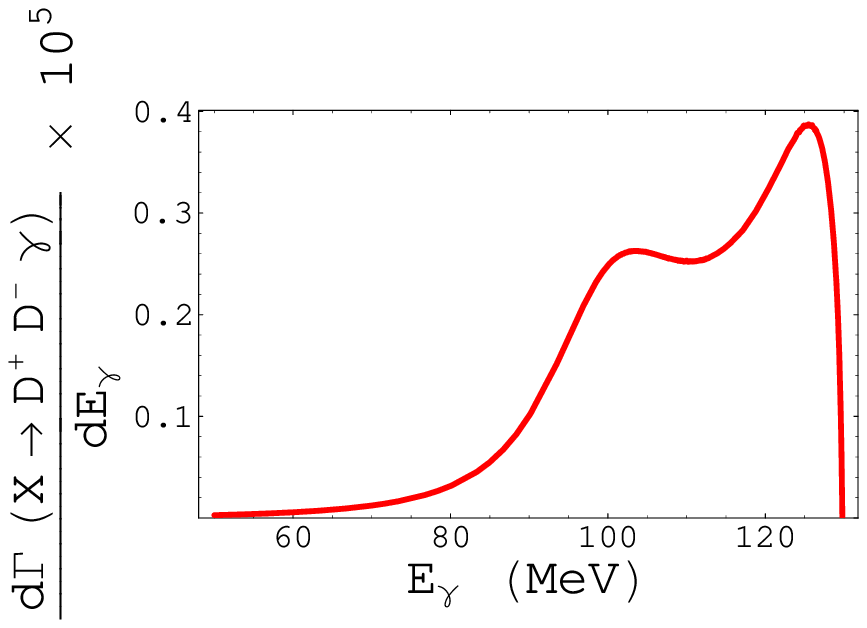} \hspace{0.35cm}
 \includegraphics[scale=0.5]{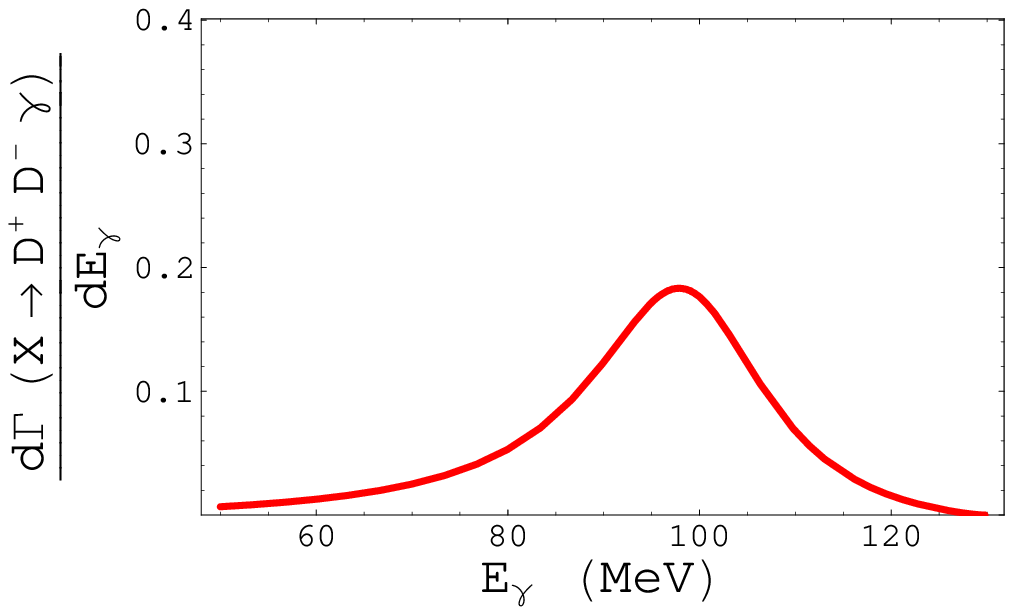}
\end{center}
\caption{\footnotesize{Photon spectrum (in arbitrary units) in $X
\to D^0 \bar D^0 \gamma$ (top) and  $X \to D^+ D^- \gamma$ (bottom)
decays  for values of the hadronic parameter $c/\hat g_1=1$ (left)
and $c/\hat g_1=300$ (right).}}\label{fig:spectra} \vspace*{1cm}
\end{figure}

The photon spectrum in radiative $X$ decays to both neutral and
charged $D$ meson pairs for two representative values of $c/ \hat
g_1$, namely $c/\hat g_1=1$ and $c/\hat g_1=300$, is depicted in
fig.\ref{fig:spectra}.
For small values of the parameter $c/\hat g_1$, i.e. in the
condition where the intermediate $D^*$ dominates the decay
amplitude, the photon spectrum in the $D^0 \bar D^0 \gamma$ mode
essentially coincides with the line corresponding to the $D^*$ decay
at $E_\gamma \simeq 139$ MeV and width determined by the $D^*$
width. The narrow peak is different from the line shape expected in
a molecular description, which is related to the wave function of
the two heavy mesons bound in the $X(3872)$,  in particular  to the
binding energy of the system, being broader for larger binding
energy. On the other hand, the photon spectrum in the charged $D^+
D^- \gamma$ mode is broader, with a peak at  $E_\gamma \simeq 125$
MeV, the total  $X \to D^+  D^- \gamma$ rate being severely
suppressed with respect to the   $X \to D^0  \bar D^0 \gamma$ one.

At the opposite side of the $c/\hat g_1$ range,  where $\psi(3770)$
gives a large contribution to the radiative amplitude, a peak at
$E_\gamma \simeq 100$ MeV appears both in neutral and charged $D$
meson modes, in the first case together with  the structure at
$E_\gamma \simeq 139$ MeV. This spectrum was described also in
\cite{voloshin1}, where in this case the radiative decay was
interpreted as deriving from the $\bar c c$ core of  $X(3872)$. In
this range of parameters  the ratio of the $X \to D^+  D^- \gamma$
to  $X \to D^0  \bar D^0 \gamma$ rates reaches the largest value.

The experimental determination of the photon spectrum of the type
depicted in fig.\ref{fig:spectra}, together with the measurement of
the $X \to D \bar D \gamma$ widths is a challenging task.
Nevertheless, this  measurement is   important to  shed light on the
structure of $X(3872)$. Other methods to distinguish between a
molecular and a quarkonium description have been proposed, and
concern the pionic $X$ decays to $\chi_{cJ}$
\cite{Dubynskiy:2007tj}.

\section*{Conclusions}
\label{sec:3}

I have summarized the features of $X(3872)$, discussing in
particular possible tests of the molecular interpretation of this
state. I have focused the attention on the radiative decays $X\to
D\bar D\g$, showing that the suppression of the charged channel with
respect to the neutral one is not a distinctive feature of the
molecular picture. However, in this decay mode the photon spectrum
is expected to be different in the $c\bar c$ and molecular
description, so that its experimental investigation could be useful
to shed more light on this puzzling hadron.

\section*{Acknowledgements}
  I thank the conference organizers and convenors.
  I also acknowledge P.~Colangelo and F.~De~Fazio for collaboration.
  This work has been supported in part by the EU Contract  No. MRTN-CT-2006-035482, "FLAVIAnet".

\end{document}